\documentclass[aps,floatfix,superscriptaddress,notitlepage,nofootinbib]{revtex4-1}
\usepackage{amsmath,amssymb,amsfonts,graphics,graphicx,dcolumn,bm,enumerate}
\usepackage{comment,natbib,appendix}
\usepackage{multirow,color}
\usepackage{chngpage}
\usepackage{afterpage}
\usepackage{xcolor}
\usepackage{amsthm}
\usepackage{natbib}
\usepackage{hyperref}
\usepackage[margin=1.2in]{geometry}
\usepackage{epstopdf}
\usepackage{float}
\usepackage{amsmath}
\usepackage{adjustbox}
\usepackage{booktabs}
\usepackage{cleveref}
\usepackage{hyperref}

\newcommand{\cmp}
{\affiliation{Saha Institute of Nuclear Physics, Kolkata 700064, India.}}
\newcommand{\isi}
{\affiliation{Economic Research Unit, Indian Statistical Institute, Kolkata 700108, India.}}
\newcommand{\raghunathpur}
{\affiliation{Department of Physics, Raghunathpur College, Raghunathpur, Purulia 723133, India.}}

\begin{document}
%\title{Relations Between the Inequality Indices Gini, Hoover and Kolkata: Theory and Data Analysis of IRS Income, Box Office Income, Citations of Nobel Laureates}
\title{Relations Between the Inequality Indices Gini, Pietra and Kolkata: Theory and Data Analysis}

\author{Asim Ghosh}
\email[Email: ]{asimghosh066@gmail.com}
\raghunathpur
 
 \author{Bikas K. Chakrabarti}%
 \email[Email: ]{bikask.chakrabarti@saha.ac.in}
 \isi  \cmp 
 
\begin{abstract}
We study relations between three inequality indices, namely the Gini
($g$), Pietra ($p$) and Kolkata ($k$) introduced in 1912, 1915 and 2014
respectively and all are  derived from the Lorenz function $L(x)$
introduced in 1905. The Kolkata index (which corresponds to a fixed point
of the complementary Lorenz function $L_c(x) \equiv 1-L(x)$)   gives the
fraction $k$ of wealth  possessed by the richest $1-k$ fraction of people
($k$ = 0.8 corresponds to Pareto's 80-20 law from 1896). We show rigorously that while the Pietra index value $p$ should be
greater than or equal to $2k-1$, the Robin Hood index should strictly
be equal to the excess wealth fraction $2k-1$ possessed by the richest
$1-k$ fraction of people. Our numerical
data analysis for US IRS Income data (1983-2022), Bollywood (India) movie
income data (1999-2024) and the citation inequalities across the
publications by forty Nobel Laureates (2020-2025) in Economics, Physics,
Chemistry and Medicine
clearly show that $p/(2k-1)$ is always greater than unity but the
deviation is never more than five percent. Assuming some simple analytic
form for the Lorenz function, we also derived the relations $k = (1/2) +
(3/8)g$ for small $g$ values and $p/g = 3/4$. However, by considering the Lorenz function appropriate for
the generalized Pareto power law distribution, we obtain an
extended range ($1_+ < p/g \leq e/2$) that captures most of the empirical data.
\end{abstract}

\maketitle

\section{Introduction}
\begin{figure}[h]
    \centering
    \includegraphics[width=1.0\linewidth]{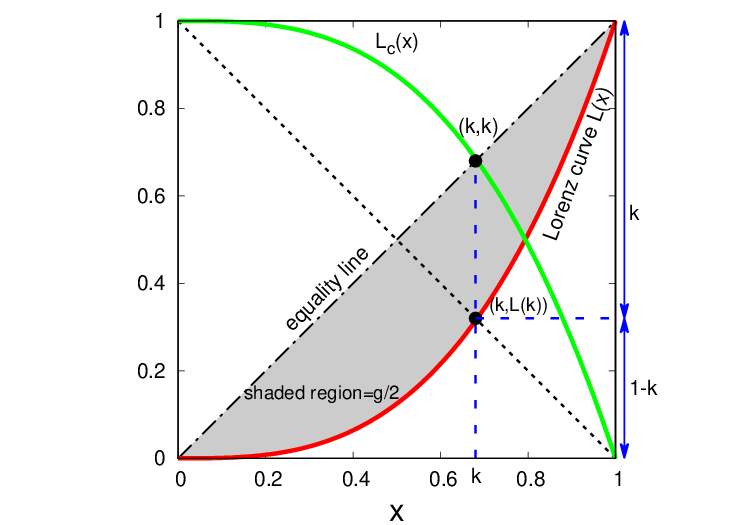}
    \caption{The Lorenz curve or function ($L(x)$, in red) shows the proportion of total wealth owned by a fraction ($x$) of people in ascending order of wealth. The black dotted-dashed line represents a scenario of perfect equality in which everyone possesses the same amount of wealth. The shaded region has area $g/2$. Consequently, the Gini index is defined as twice of this area because the area under the equality line is 1/2.  The Pietra index $p$ is given by the length of the maximum vertical distance of the Lorenz curve from the equality line and is indicated here by the solid vertical line in black. The complementary Lorenz function ($L_c(x) \equiv 1 - L (x)$) is shown in green. The Kolkata index
($k$) is determined by the fixed point  of the complementary Lorenz curve: $L_c (k) = k$ or
$L(k) = 1 - k$.  Geometrically, it gives the point at which the Lorenz curve intersects the
diagonal line perpendicular to the equality line, and it gives the fraction
$k$ of wealth that is possessed by the richest $1 - k$ fraction of
people. As such, $k = 0.8$ corresponds to Pareto's 80-20 law.}
    \label{fig:lorenz}
\end{figure}

Quantitative measurements of economic inequalities, specifically the
income or wealth inequalities, in different societies or countries started
with the introduction of Lorenz function \cite{Lorenz1905} in 1905. The Lorenz function
or curve is represented graphically  by plotting the cumulative fraction
of the income or wealth ($L(x)$) earned or possessed by the $x$ fraction
of the poorest individuals, when the population of the society is arranged
in the ascending order of their income (from now on income will mean
wealth of any kind of agent in countries, societies, institutions etc.;
see fig. \ref{fig:lorenz}). When everyone earns the same, the Lorenz curve becomes the
dotted-dashed (black) diagonal in fig. \ref{fig:lorenz}. Because of the inequalities among the
agents (and their ordering from the poorest to the richest), the typical
shape of the Lorenz curve will have the form of the red line (with $L(0) =
0$ and $L(1) = 1$). Although $L(x)$ contains all the information of
inequality, typical values of the inequality indices help summarizing or
characterizing the nature of inequality in society. One of the oldest
and still most popular one is the Gini index \cite{Gini1912} ($g$) given by the
normalized area between the equality line and the Lorenz curve (see fig. \ref{fig:lorenz}). The Pietra index \cite{Pietra15} ($p$), also known as \cite{Allen22} Robin Hood index or Hoover index or Schulz
index, is given by the maximum vertical distance between the equality line
and the Lorenz curve (see fig. \ref{fig:lorenz}). These index values are measured using fractions here (equivalent to the respective percentage values commonly used in the economics literature).

%of the total population’s income that would have to be
%redistributed to make everyone’s income equal.

%The Gini index ($g$) is calculated from the area between the Lorenz curve and the equality line (shaded region), normalized by the total area (= 1/2) under the equality line.

In 2005 Hirsch noted  \cite{Hirsch05} that the success in citations of the papers by
the scientists could be quantified by an index $h$ (known as Hirsch index
today) corresponding to the fixed point of the nonlinear (Ziff-like) citation function ($f(c)$) giving the number
$f$ of papers having citation $c$ : $h = f(h)$. Soon this index became a very popular one to quantify
the success inequalities among the scientists. Inspired by that, and also
noting that  the nonlinear Lorenz function $L(x)$ has trivial fixed points at $x = 0$
and $x = 1$, we defined \cite{Ghosh14} the complementary Lorenz function $L_c(x) \equiv 1
- L(x)$, which has a nontrivial fixed point at $k$ such that $L_c(k) = k$,
where $k \geq 1/2$ (see fig. \ref{fig:lorenz}). This $k$ is called the Kolkata index (see
also \cite{Banerjee20,Banerjee23}) to epitomize the extreme economic inequality in this Indian
city. 

It is worth mentioning here that another another Lorenz based
inequality measure, namely the Zanardi index, which has been discussed
recently \cite{Stiglitz19,Landini25}. The Zanardi index value is expected  to have some
relationship with  the Kolkata index value since they both rely on the position
of the same point on the Lorenz curve (where the complementary
equality line crosses the Lorenz curve), though no exact relationship
between these indices could be found.

A simple geometric argument for the Lorenz function then
suggests that $1-k$ fraction of the rich people possess $k$ fraction of
the total wealth. As such it generalizes then the Pareto’s 80-20 law \cite{Pareto1896} where $k = 0.80$.
It was later observed (see e.g., \cite{Banerjee23,Manna22}) that in physical systems with
self-organized competitive dynamics among the constituent degrees of
freedom the inequality levels (in all cluster or avalanche size
distributions) tend
to grow towards a universal level $g = k \simeq 0.87$ (somewhat above the
Pareto value $k = 0.80$) and remains thereafter in the self-organized
critical state of the system. Later, assuming \cite{Joseph22} a minimal polynomial
expansion of the Lorenz function $L(x)$, we could derive a simple linear
 relationship between the Gini and Kolkata indices growing up to $g =
k = 0.80$. This linear relationship of Kolkata ($k$) with Gini ($g$)
indices agreed generally with our data analysis (see \cite{Banerjee23B}) for lower
values of $g$.

\begin{figure}[H]
    \centering
    \includegraphics[width=1.0\linewidth]{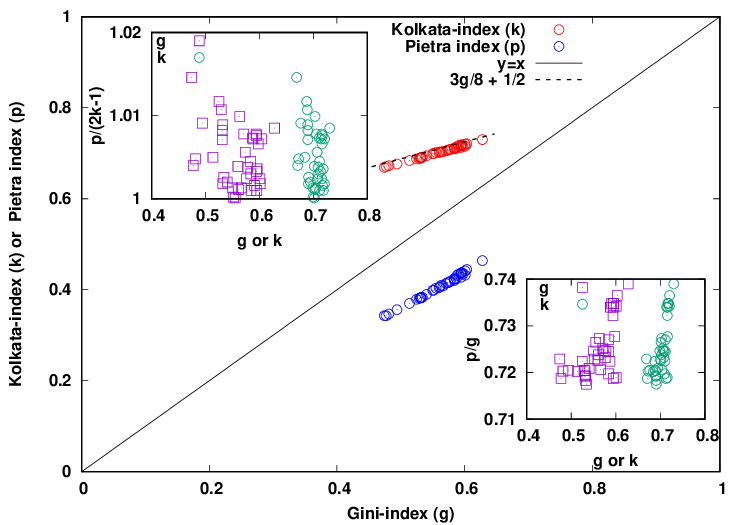}
    \caption{Growths of Pietra ($p$) and Kolkata ($k$) indices against Gini ($g$) index
values for US economy 
income data (IRS data \cite{irs,Yakovenko22} for the period 1982-2022). The values
of $k$ and $g$ are growing with time, because of increasing rate of
withdrawal of public welfare programs. The upper left inset,
showing the
values of $p/(2k - 1)$ against the years, indicate a value higher than
unity as predicted by a theoretical argument (relation (\ref{eqn3}). The
lower right inset shows the values of $p/g$  against years and show values
considerably different from 3/4, as obtained from the theoretical relation
(\ref{eqn5}) (obtained an additional assumption (\ref{eqn4}) of the minimal form of the
Lorenz function).  Details of the values of the inequality indices and
of their relations are given in Table I of the Appendix.}
    \label{fig2:income}
\end{figure}

\begin{figure}[h!]
    \centering
    \includegraphics[width=1.0\linewidth]{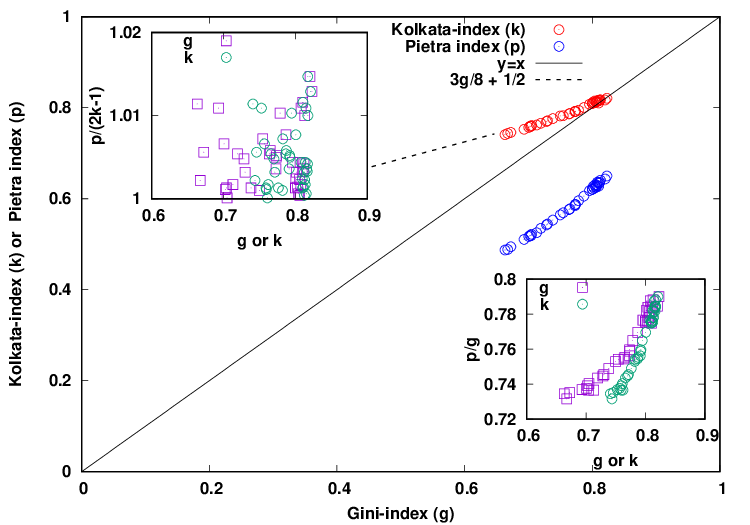}
    \caption{Growths of Pietra ($p$) and Kolkata ($k$) indices against Gini ($g$) index
values for US economy income tax return data (IRS data \cite{irs,Yakovenko22} for the
period 1982-2022). The values of $k$ in the tax data (which can be
argued to represent the prevailing inequality status better) seem to have
grown a little beyond the Pareto value ($k = 0.80$) with increasing
withdrawal of public welfare programs. The upper left  and lower right insets, showing the values of $p/(2k-1)$ against $g$ and $k$ respectively, indicate values  slightly higher
than unity (as predicted by a theoretical argument; see relation (\ref{eqn3}). The lower right inset shows the values of $p/g$
against years and show values considerably different from 3/4, as
obtained from the theoretical relation (\ref{eqn5}) (obtained an additional
assumption (\ref{eqn4}) of the minimal form of the Lorenz function). Details of
the  values of the inequality indices and
of their relations are given in Table II of the Appendix.}
\label{fig3:tax}
\end{figure}

\section{Theoretical analysis of Inequality indices}
We will first explore analytically if there are any relations among these
three indices, namely  Gini ($g$), Pietra ($p$) and Kolkata ($k$). To
begin with, we note that since by construction the Kolkata index ($k$) is
given by the fraction of the total wealth possessed by the $ 1- k$
fraction of richest people ($k > 1/2$; $k = 1/2$ corresponds to equality),
they  possess precisely an extra amount of $2k-1$ fraction of total
wealth.  The Pietra index $p$ , if strictly
interpreted as  the Robin Hood index \cite{Allen22}, should equal to this excess  amount  equal to $2k - 1$.

%{\color{blue}
On the other hand, the Pietra index $p$ is defined as the maximum vertical distance between the
Lorenz curve $L(x)$ and the line of perfect equality:
\begin{equation}
 p = \max_x \left(x - L(x)\right). 
 \label{eqn1}
\end{equation}
%$$  p = \max_x \left(x - L(x)\right). \eqno               $$
Evaluating the vertical distance between the equality line and the Lorenz
curve at $x=k$, we obtain
\begin{equation}
\left. x - L(x) \right|_{x=k}
= k - (1-k)
= 2k - 1 .
\label{eqn2}
\end{equation}
Since the Pietra index is the maximum value of this vertical distance over all $x$, it follows that
\begin{equation}
p \ge 2k - 1 . 
\label{eqn3}
\end{equation}
%}

\subsection{Analytical relation connecting $g$, $k$ and $p$ using a quadratic form of Lorenz function}

In order to find theoretically the other plausible relations among those
indices, let us proceed with a (Landau-like; see e.g., \cite{Joseph22}) minimal
polynomial expansion of the Lorenz function:

\begin{equation}         L(x) = Ax + Bx^2,                
\label{eqn4}           
\end{equation}

\noindent where $A > 0$ and $B > 0$, such that $L(x)$ becomes a
monotonically increasing function of $x$ and $A + B = 1$ to ensure $L(x) =
1$ for $x = 1$ (also ensuring $L(x) = 0$ at $x = 0$). One can then express
the Gini index through

\begin{equation*}  
g = 1- 2\int_0^1 L(x) dx = 1 - A -(2/3) B = B/3.            
\end{equation*}

\noindent The Pietra index $p$ is given by the maximum value of the
difference between $x$ (equality line) and $L(x)$ (Lorenz curve) for $x$ varying from 0 to 1:

\begin{equation*} p = max [x - L(x)] = max [B(x - x^2] = B/4,                  
\end{equation*}
\noindent  giving,
\begin{equation}        p   = (3/4) g.                        
\label{eqn5}
\end{equation}
\noindent Also noting that the Kolkata index $k$ is given by the  fixed point equation $L_c(k)=k$ or the complementary Lorenz function $L_c(x) \equiv 1-L(x)$, giving (see also \cite{Joseph22,Banerjee23B})

\begin{equation} k    = 1/2 + (3/8)g,                            
\label{eqn6}
\end{equation}
\noindent for $g \rightarrow 0$.

\subsection{Analytical Relation connecting $g$ and $p$ using the Pareto form
of Lorenz function}
\begin{figure}[h!]
\centering
\includegraphics[width=1.0\linewidth]{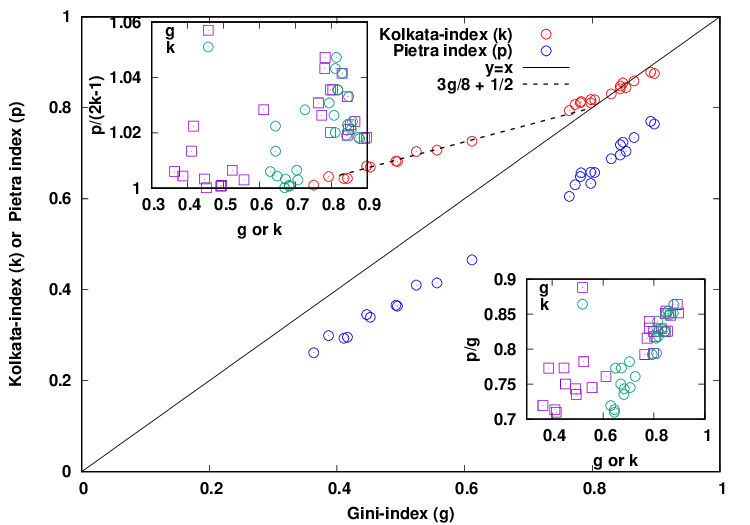}
\caption{Growths of Pietra ($p$) and Kolkata ($k$) indices against Gini ($g$) index
values for movie income inequalities (Bolloywood India data \cite{13bbollywood} for the
period 1999-2024).  The values of $k$ in the tax data (which can be
argued to represent the prevailing inequality status better) seems to have
grown a little beyond the Pareto value ($k$ = 0.80) with increasing
withdrawal of public welfare programs. The upper left inset, showing the
values of $p/(2k - 1)$ against the years, indicate values higher than
unity as predicted by a theoretical argument (relation (\ref{eqn3}). The
lower right inset shows the values of $p/g$  against years and show values
considerably different from 3/4, as obtained from theoretical relation (\ref{eqn5})
(obtained an additional assumption (\ref{eqn4}) of the minimal form of the Lorenz
function). Details of the values of the inequality indices and of
their relations are given in Table III of the Appendix.}
\label{fig4:movie}
\end{figure}

\begin{figure}[h!]
\centering
\includegraphics[width=1.0\linewidth]{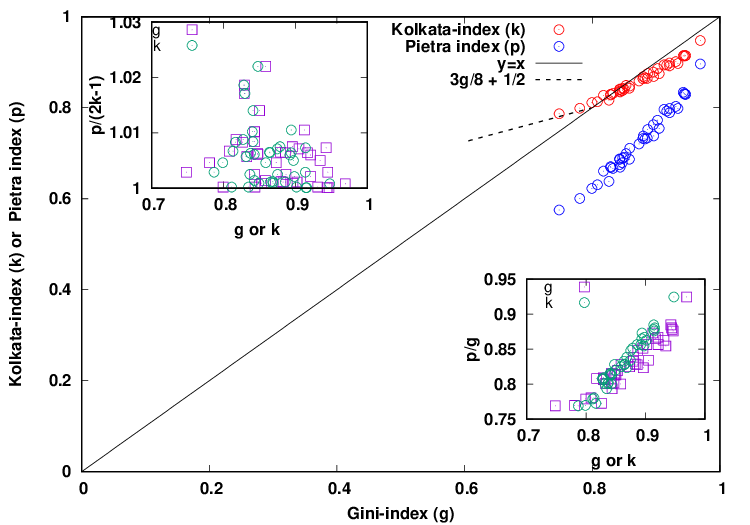}
\caption{Growths of Pietra ($p$) and Kolkata ($k$) indices against Gini ($g$) index
values for the citation inequalities among the papers published by
different Nobel Laureates in economics, physics, chemistry and medicine
during the last six years (Google Scholar open-access data \cite{google-scholar} for
scientists, each having their own home page with verified e-mail address
and more than 100 papers). The values of $k$ (and also of $g$ for the inequalities in citation
distributions across the publications of individual Nobel laureates) have
grown, because of extreme competition among the scientists,  quite beyond
the Pareto value ($k$ = 0.80, and closer to self-organized critical system
value \cite{Manna22,Banerjee23B}) because of any public welfare kind of support system (for
the producers) and the market being quite competitive. The upper left  and
lower right insets, showing the values of $p/(2k - 1)$ against $g$ and $k$
respectively, indicate values  slightly higher than unity (as predicted by
a theoretical argument; see relation (\ref{eqn3}). The lower right inset shows the values of $p/g$
against years and show values considerably different from 3/4, as
obtained from the theoretical relation (\ref{eqn5}) (obtained an additional
assumption (\ref{eqn4}) of the minimal form of the Lorenz function). Details of
the inequality indices and of their relations are given in Table IV of the
Appendix.}
\label{fig5:nobel}
\end{figure}

\begin{figure}
    \centering
    \includegraphics[width=0.9\linewidth]{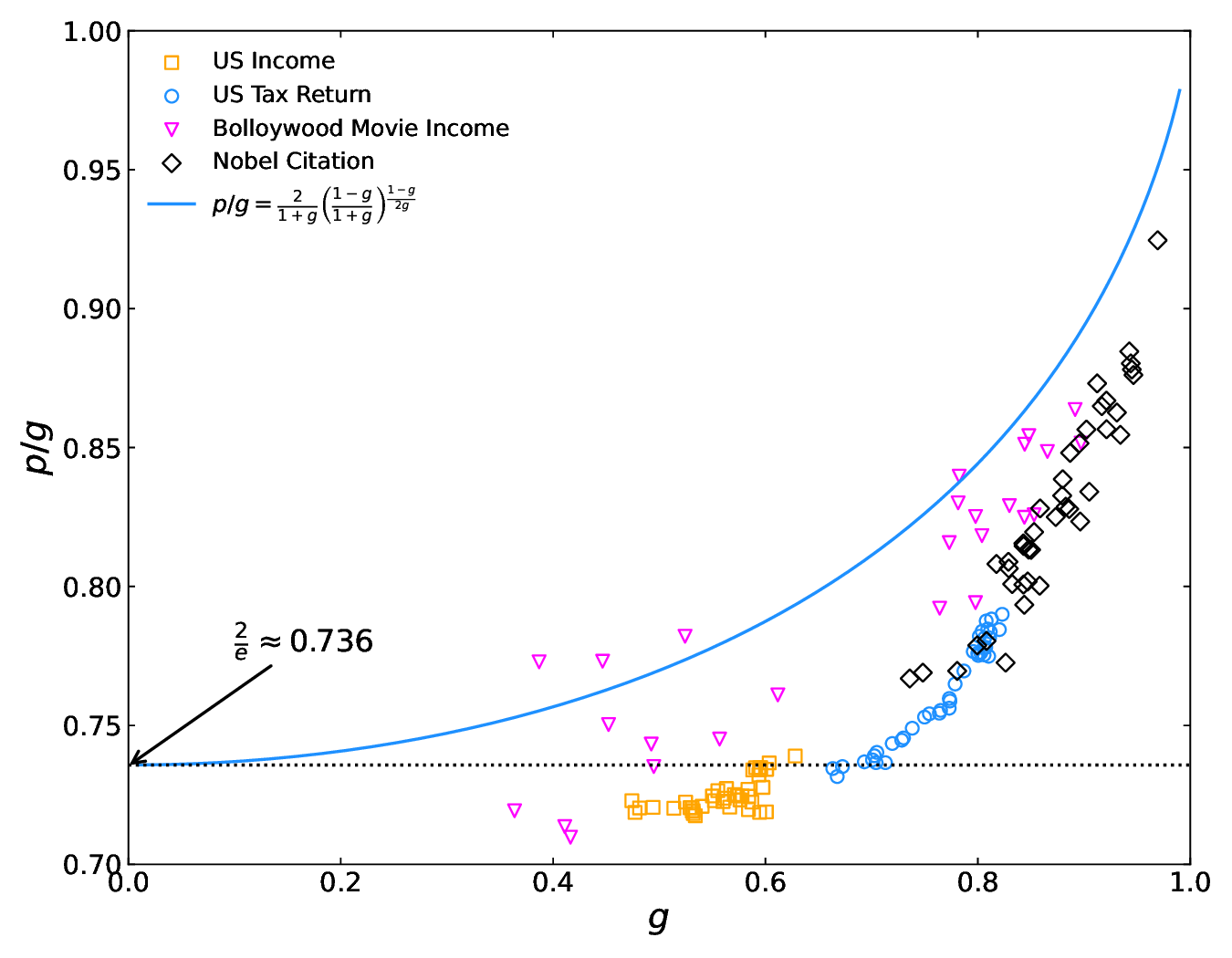}
    \caption{We compare the theoretical results of eq. (\ref{eqn11}) with the
numerical data analysis for the inequalities in the US economy income as
well as income tax return data (IRS data \cite{irs,Yakovenko22} for the period
1982-2022), for inequalities in movie income data (Bollywood India
\cite{13bbollywood} for the period 1999-2024)  and for the citation inequalities
among the papers published by different Nobel Laureates in economics,
physics, chemistry and medicine during the last six years (Google
Scholar open-access data \cite{google-scholar} for scientists, each having their own
home page with ``verified e-mail address'' and more than 100 published
papers).}
    \label{fig:6}
\end{figure}

To complement the quadratic approximation discussed above, we consider
the generalized Lorenz function appropriate of the Pareto distribution (see, for example, \cite{Banerjee23B,biswas23}):
\begin{equation}
L(x)=1-(1-x)^{1-\frac{1}{n}}, \qquad n>1.
\end{equation}
Let us define $D(x)=x-L(x)=x-1+(1-x)^{\alpha}$, where $\alpha=1-1/n$. If we take a derivative of $D(x)$ and set that to zero, we will get the maximizing point at
    \begin{align*}
    x^{*}&=1-(1-\frac{1}{n})^{n} ~.
    \end{align*} 
After putting the value of $x^{*}$ in $D(x)$ and after some simplification we will get the Pietra index 
\begin{equation}
p=\frac{1}{n}(1-\frac{1}{n})^{n-1}.
\end{equation} 
And 
\begin{align*}
g
&=
\frac{\displaystyle \frac{1}{2}-\int_{0}^{1}L(x)\,dx}{\displaystyle \frac{1}{2}} = 1-2\int_{0}^{1}\left[1-(1-x)^{\alpha}\right]dx,
\end{align*}
giving
\begin{align}
g 
&=
\frac{1-\alpha}{1+\alpha}= \frac{1}{2n-1}.
\end{align}
Hence
\begin{equation}
\frac{p}{g}=\frac{2n-1}{n}(1-\frac{1}{n})^{n-1}~.
\end{equation} 
For $n=2$, this gives $p/g=3/4$,  which is the result of
quadratic approximation of $L(x)$ (see Eq.~(5)). As $n$ increases, the ratio
$p/g$ decreases smoothly and approaches the asymptotic value
$2/e\approx0.736$ as $n\rightarrow\infty$. Thus, $p/g$ varies continuously
from $1_+$ (as $n\rightarrow1_+$) to $2/e\approx0.736$ (as
$n\rightarrow\infty$), consistent with the empirical values shown in the
insets of \cref{fig2:income,fig3:tax,fig4:movie,fig5:nobel}.  It may be noted that by expressing $n$ in terms of $g$ from
equation (9) and putting that in equation (10), we get
\begin{equation} 
p/g=\frac{2}{1+g}\left(\frac{1-g}{1+g}\right)^{\frac{1-g}{2g}}                     \label{eqn11}
\end{equation}
\noindent In Fig. \ref{fig:6}, we compare this theoretically expected variation
of $p/g$ as a function of $g$ with those obtained from our numerical
data analysis discussed in the next section.

\section{Numerical Data Analysis}
We will now analyze the extensive data sets from various income data sources
(US IRS for the forty year period 1983 to 2022 \cite{irs,Yakovenko22}, Bollywood Box Office
income data for the period 1999 to 2024 \cite{13bbollywood}), comparing how the above
relations (\ref{eqn3}), (\ref{eqn5}) and (\ref{eqn6}) fit the results of
our data analysis. We will analyze
extensive citation data (from Google Scholar \cite{google-scholar}) of the prolific  Nobel
laureate research scientists (physicists, chemists, medicinal scientists
and economists) for the years 2020 to 2025.

\section{Summary and Conclusion}
As we discussed, all three inequality indices Gini ($g$), Pietra
($p$) and Kolkata ($k$) are derived from the Lorenz function $L(x)$
(see fig. \ref{fig:lorenz}). $g/2$ is given by the area between the equality line and
the Lorenz curve. $p$ is given by the maximum value of $x-L(x)$ and is
interpreted as a measure of the fraction of money the `rich'
people should transfer to the `poor' people (neither rich or poor are specifically 
defined here), so that a redistribution would bring equality (Robin
Hood index \cite{Allen22}). $k$ is given by the fixed point of the Complementary
Lorenz Function $L_c(x) \equiv 1 - L(x)$: $L_c(k) = k$. Geometrically
(see fig. \ref{fig:lorenz}) it gives the fraction $k$ of  wealth (citations)
possessed by clearly defined $1 - k$ fraction of rich people (papers). We therefore
argued the excess wealth of the rich is precisely equal to $2k - 1$,
which if they transfer to the poor $k$ fraction of people, then (on
redistribution) equality would be achieved. As such, the Robin Hood index should be equal to $2k
-1$, while the value of Pietra Index $p$, given by the maximum
vertical distance of the Lorenz curve from the equality line is given
by the relation (\ref{eqn3}), with $2k-1$ as its  minimum value.

As we see from numerical data analysis (see Appendices I-IV) and
upper-left insets of \cref{fig2:income,fig3:tax,fig4:movie,fig5:nobel} the $p$ is slightly higher (at most 5\% off) from the value ($2k-1$ of the Robin Hood index), as given by the relation (\ref{eqn3}). Next we
assumed a minimal analytic and polynomial form \ref{eqn4}  for the Lorenz function
$L(x)$ and calculated $g$ and $p$ from there, giving $p = (3/4)g$
(relation (\ref{eqn5})). As we see from the tables I-IV and  \cref{fig2:income,fig3:tax,fig4:movie,fig5:nobel}, there
are large deviations from relation (\ref{eqn5}). Finally, solving for $k$ from
(\ref{eqn4}) and (\ref{eqn5}), in the small $g$ limit, we got $k =  (1/2) + (3/8)g$
(relation (\ref{eqn6}); see section II.A). Numerical data analysis for the validity of this
relation is indicated in each of the main parts and bottom right insets of  \cref{fig2:income,fig3:tax,fig4:movie,fig5:nobel}. In section II.B, we incorporate the analytical
results for $p/g$ by considering the Lorenz function appropriate for
the generalized Pareto (single power) law distribution. This yields an
extended range for the ratio $p/g$, going smoothly from $1_+$ at $n=1_+$ to $2/e\approx0.736$ when $n\rightarrow\infty$. This extended range accumulated most of empirical data shown in the insets of \cref{fig2:income,fig3:tax,fig4:movie,fig5:nobel}.  It may be noted that this analytical calculation  could  not be calculated  for
$k$ index  value. Therefore any extended version of relation (\ref{eqn6}),  for general values of  $n$,  could not be obtained.

We studied the relations between three inequality indices, namely the
Gini ($g$) \cite{Gini1912}, Pietra ($p$) \cite{Pietra15} and Kolkata ($k$) \cite{Ghosh14,Banerjee20,Banerjee23} introduced in
1912, 1915 and 2014 respectively and all are derived from the Lorenz
function \cite{Lorenz1905} $L(x)$ introduced in 1905. We showed rigorously that while the Pietra index $p$ should have the minimum value $2k-1$
(relation (\ref{eqn3}), the Robin Hood index value will be given  the fraction $2k -
1$ of excess wealth possessed by the richest $1-k$ fraction of
people. Assuming
some simple analytic form (\ref{eqn4}) in section II.A for the Lorenz function (cf. \cite{Joseph22}), we also
derived here the relations (\ref{eqn5}) $p/g = 3/4$ and (\ref{eqn6}) $k = (1/2) + (3/8)g$ for small values of $g$. In section IIB, by considering the Lorenz function
appropriate for the generalized Pareto (single power) law distribution, we get an
extended range ($1_+ < p/g \leq e/2$) for the ratio ($p/g$); which accumulated most of empirical data shown in the insets of \cref{fig2:income,fig3:tax,fig4:movie,fig5:nobel}. 
Our numerical
data analysis results for US IRS Income data (1983-2022; see Tables I, II
of the Appendix, and  \cref{fig2:income,fig3:tax}), Bollywood (India) movie income data
(1999-2024; see Table III of the Appendix and fig. \ref{fig4:movie}) and the citation
inequalities across the publications by forty Nobel Laureates (2020-2025;
see Table IV of the Appendix and fig. \ref{fig5:nobel}) in Economics, Physics, Chemistry
and Medicine clearly shows that $p/(2k -1)$ is always slightly greater
than unity and the deviation is never more than five percent. Assuming
some simple analytic form (\ref{eqn4}) for the Lorenz function (cf. \cite{Joseph22}), we also
derived here the relations (\ref{eqn5}) $p/g = 3/4$ and (\ref{eqn6}) $k = (1/2) + (3/8)g$ for small values of $g$. Indeed, we think these  indications (from  \cref{fig2:income,fig3:tax,fig4:movie,fig5:nobel})
of sustained hovering of the social index values  near $g = k \simeq
0.86$ suggest self-organised criticality (see e.g., \cite{Manna22,biswas23})  for a self-tuned socio-dynamical system.

\begin{acknowledgments}
 We are indebted to an anonymous reviewer for suggesting the inclusion of Section IIB.
\end{acknowledgments}

\vskip 10 cm

\begin{appendices}
\section*{Appendix: Detailed numerical estimates of inequality indices from data analysis}
%\newpage
\begin{table}[h]
\centering
\caption{Gini-index ($g$), Pietra-index ($p$) and Kolkata-index ($k$) 
 values for the  income inequalities from  U.S. IRS Income
Data (1983–2022)  collected from ref. \cite{irs,Yakovenko22}. For each
year, the total income collections of individuals are analyzed.}
%\begin{adjustbox}{width=0.6\textwidth}
\begin{tabular}{|l|c|c|c|c|c|c|}
\hline
%\textbf{Year} & \textbf{$g$-index} & \textbf{$k$-index} & \textbf{$p$-index} & $p/(2k-1)$ & p/g & 2k-1/g  \\ \hline
\toprule
\adjustbox{angle=90}{Year} &
\adjustbox{angle=90}{Gini index ($g$)} &
\adjustbox{angle=90}{Pietra index ($p$)} &
\adjustbox{angle=90}{Kolkata index ($k$)}&
\adjustbox{angle=90}{$p/(2k-1)$} &
\adjustbox{angle=90}{$p/g$} &
\adjustbox{angle=90}{$(2k-1)/g$}\\
\midrule
\hline
1983 & 0.47418 & 0.34280 & 0.66894 & 1.0146 & 0.7229 & 0.7126\\ \hline 
1984 & 0.47711 & 0.34289 & 0.67076 & 1.0040 & 0.7187 & 0.7158\\ \hline 
1985 & 0.48136 & 0.34672 & 0.67253 & 1.0048 & 0.7203 & 0.7168\\ \hline 
1986 & 0.49412 & 0.35602 & 0.67640 & 1.0091 & 0.7205 & 0.7140\\ \hline 
1987 & 0.51365 & 0.36991 & 0.68403 & 1.0050 & 0.7202 & 0.7166\\ \hline 
1988 & 0.53394 & 0.38312 & 0.69106 & 1.0026 & 0.7175 & 0.7157\\ \hline 
1989 & 0.53094 & 0.38250 & 0.68970 & 1.0082 & 0.7204 & 0.7146\\ \hline 
1990 & 0.52898 & 0.38110 & 0.68854 & 1.0107 & 0.7204 & 0.7128\\ \hline 
1991 & 0.52472 & 0.37906 & 0.68734 & 1.0117 & 0.7224 & 0.7141\\ \hline 
1992 & 0.53185 & 0.38256 & 0.68960 & 1.0089 & 0.7193 & 0.7130\\ \hline 
1993 & 0.53100 & 0.38191 & 0.68960 & 1.0071 & 0.7192 & 0.7141\\ \hline 
1994 & 0.53198 & 0.38212 & 0.69071 & 1.0018 & 0.7183 & 0.7170\\ \hline 
1995 & 0.54026 & 0.38946 & 0.69434 & 1.0020 & 0.7209 & 0.7194\\ \hline 
1996 & 0.55188 & 0.39894 & 0.69943 & 1.0002 & 0.7229 & 0.7227\\ \hline 
1997 & 0.55998 & 0.40463 & 0.70153 & 1.0039 & 0.7226 & 0.7198\\ \hline 
1998 & 0.56631 & 0.40807 & 0.70376 & 1.0013 & 0.7206 & 0.7196\\ \hline 
1999 & 0.57604 & 0.41659 & 0.70780 & 1.0024 & 0.7232 & 0.7215\\ \hline 
2000 & 0.58418 & 0.42321 & 0.71140 & 1.0010 & 0.7245 & 0.7237\\ \hline 
2001 & 0.56116 & 0.40616 & 0.70285 & 1.0011 & 0.7238 & 0.7230\\ \hline 
2002 & 0.55009 & 0.39857 & 0.69902 & 1.0013 & 0.7246 & 0.7236\\ \hline 
2003 & 0.55508 & 0.40325 & 0.70161 & 1.0001 & 0.7265 & 0.7264\\ \hline 
2004 & 0.57073 & 0.41385 & 0.70533 & 1.0078 & 0.7251 & 0.7195\\ \hline 
2005 & 0.58713 & 0.42413 & 0.71052 & 1.0073 & 0.7224 & 0.7171\\ \hline 
2006 & 0.59445 & 0.42721 & 0.71290 & 1.0033 & 0.7187 & 0.7163\\ \hline 
2007 & 0.60078 & 0.43191 & 0.71543 & 1.0024 & 0.7189 & 0.7172\\ \hline 
2008 & 0.58395 & 0.42027 & 0.70950 & 1.0030 & 0.7197 & 0.7175\\ \hline 
2009 & 0.56321 & 0.40960 & 0.70280 & 1.0099 & 0.7273 & 0.7202\\ \hline 
2010 & 0.57349 & 0.41558 & 0.70664 & 1.0056 & 0.7247 & 0.7206\\ \hline 
2011 & 0.57799 & 0.41882 & 0.70862 & 1.0038 & 0.7246 & 0.7219\\ \hline 
2012 & 0.59759 & 0.43488 & 0.71601 & 1.0066 & 0.7277 & 0.7229\\ \hline 
2013 & 0.58345 & 0.42418 & 0.71113 & 1.0045 & 0.7270 & 0.7237\\ \hline 
2014 & 0.59374 & 0.43475 & 0.71569 & 1.0078 & 0.7322 & 0.7265\\ \hline 
2015 & 0.59442 & 0.43629 & 0.71649 & 1.0076 & 0.7340 & 0.7284\\ \hline 
2016 & 0.58799 & 0.43158 & 0.71424 & 1.0072 & 0.7340 & 0.7287\\ \hline 
2017 & 0.59553 & 0.43759 & 0.71800 & 1.0036 & 0.7348 & 0.7321\\ \hline 
2018 & 0.59473 & 0.43703 & 0.71829 & 1.0010 & 0.7348 & 0.7341\\ \hline 
2019 & 0.59072 & 0.43407 & 0.71669 & 1.0016 & 0.7348 & 0.7336\\ \hline 
2020 & 0.60100 & 0.44124 & 0.72022 & 1.0018 & 0.7342 & 0.7328\\ \hline 
2021 & 0.62792 & 0.46403 & 0.73005 & 1.0085 & 0.7390 & 0.7327\\ \hline 
2022 & 0.60345 & 0.44444 & 0.72064 & 1.0072 & 0.7365 & 0.7313\\ \hline 
\end{tabular}
%\end{adjustbox}
\label{tab:income}
\end{table}

\begin{table}
\centering
\caption{Gini-index ($g$),  Pietra-index ($p$)  and Kolkata-index ($k$)
values for the tax-return inequalities from U.S. IRS
Income Tax Data (1983–2022) datasets collected from ref. \cite{irs,Yakovenko22}. For each
year, the total tax collections of individuals are analyzed.}
%\begin{adjustbox}{width=0.6\textwidth}
\begin{tabular}{|l|c|c|c|c|c|c|}
\hline
%\textbf{Year} & \textbf{$g$-index} & \textbf{$k$-index} & \textbf{$p$-index} & $p/(2k-1)$ & p/g & 2k-1/g  \\ \hline
\toprule
\adjustbox{angle=90}{Year} &
\adjustbox{angle=90}{Gini index ($g$)} &
\adjustbox{angle=90}{Pietra index ($p$)} &
\adjustbox{angle=90}{Kolkata index ($k$)}&
\adjustbox{angle=90}{$p/(2k-1)$} &
\adjustbox{angle=90}{$p/g$} &
\adjustbox{angle=90}{$(2k-1)/g$}\\
\midrule
\hline
1983 & 0.66335 & 0.48721 & 0.74086 & 1.0114 & 0.7345 & 0.7262\\ \hline 
1984 & 0.66753 & 0.48835 & 0.74363 & 1.0022 & 0.7316 & 0.7299\\ \hline 
1985 & 0.67246 & 0.49438 & 0.74581 & 1.0056 & 0.7352 & 0.7311\\ \hline 
1986 & 0.69308 & 0.51074 & 0.75262 & 1.0109 & 0.7369 & 0.7290\\ \hline 
1987 & 0.70073 & 0.51685 & 0.75674 & 1.0066 & 0.7376 & 0.7328\\ \hline 
1988 & 0.71296 & 0.52519 & 0.76215 & 1.0017 & 0.7366 & 0.7354\\ \hline 
1989 & 0.70412 & 0.51864 & 0.75899 & 1.0013 & 0.7366 & 0.7356\\ \hline 
1990 & 0.70247 & 0.51917 & 0.75930 & 1.0011 & 0.7391 & 0.7383\\ \hline 
1991 & 0.70487 & 0.52181 & 0.76089 & 1.0001 & 0.7403 & 0.7402\\ \hline 
1992 & 0.71938 & 0.53487 & 0.76601 & 1.0054 & 0.7435 & 0.7396\\ \hline 
1993 & 0.72833 & 0.54241 & 0.76992 & 1.0048 & 0.7447 & 0.7412\\ \hline 
1994 & 0.72995 & 0.54421 & 0.77123 & 1.0032 & 0.7455 & 0.7431\\ \hline 
1995 & 0.73822 & 0.55293 & 0.77611 & 1.0013 & 0.7490 & 0.7480\\ \hline 
1996 & 0.74976 & 0.56455 & 0.78198 & 1.0010 & 0.7530 & 0.7522\\ \hline 
1997 & 0.75437 & 0.56893 & 0.78243 & 1.0072 & 0.7542 & 0.7488\\ \hline 
1998 & 0.76372 & 0.57613 & 0.78651 & 1.0054 & 0.7544 & 0.7503\\ \hline 
1999 & 0.77301 & 0.58454 & 0.79122 & 1.0036 & 0.7562 & 0.7535\\ \hline 
2000 & 0.77856 & 0.59549 & 0.79470 & 1.0103 & 0.7649 & 0.7570\\ \hline 
2001 & 0.76491 & 0.57785 & 0.78725 & 1.0058 & 0.7554 & 0.7511\\ \hline 
2002 & 0.77313 & 0.58737 & 0.79221 & 1.0050 & 0.7597 & 0.7559\\ \hline 
2003 & 0.77373 & 0.58709 & 0.79204 & 1.0052 & 0.7588 & 0.7549\\ \hline 
2004 & 0.78684 & 0.60559 & 0.80048 & 1.0077 & 0.7696 & 0.7638\\ \hline 
2005 & 0.79995 & 0.62076 & 0.80998 & 1.0013 & 0.7760 & 0.7750\\ \hline 
2006 & 0.80278 & 0.62305 & 0.81109 & 1.0014 & 0.7761 & 0.7750\\ \hline 
2007 & 0.80582 & 0.62481 & 0.81229 & 1.0004 & 0.7754 & 0.7751\\ \hline 
2008 & 0.80036 & 0.62046 & 0.80949 & 1.0024 & 0.7752 & 0.7734\\ \hline 
2009 & 0.80642 & 0.63018 & 0.81371 & 1.0044 & 0.7815 & 0.7780\\ \hline 
2010 & 0.81101 & 0.63382 & 0.81576 & 1.0036 & 0.7815 & 0.7787\\ \hline 
2011 & 0.79568 & 0.61789 & 0.80758 & 1.0044 & 0.7766 & 0.7731\\ \hline 
2012 & 0.80744 & 0.62826 & 0.81359 & 1.0017 & 0.7781 & 0.7768\\ \hline 
2013 & 0.80558 & 0.62664 & 0.81260 & 1.0023 & 0.7779 & 0.7761\\ \hline 
2014 & 0.81165 & 0.63613 & 0.81674 & 1.0042 & 0.7837 & 0.7805\\ \hline 
2015 & 0.80929 & 0.63505 & 0.81648 & 1.0033 & 0.7847 & 0.7821\\ \hline 
2016 & 0.80136 & 0.62672 & 0.81236 & 1.0032 & 0.7821 & 0.7796\\ \hline 
2017 & 0.80401 & 0.63025 & 0.81495 & 1.0006 & 0.7839 & 0.7834\\ \hline 
2018 & 0.81284 & 0.64080 & 0.81722 & 1.0100 & 0.7883 & 0.7805\\ \hline 
2019 & 0.80787 & 0.63630 & 0.81471 & 1.0109 & 0.7876 & 0.7791\\ \hline 
2020 & 0.82292 & 0.65008 & 0.82091 & 1.0129 & 0.7900 & 0.7799\\ \hline 
2021 & 0.82025 & 0.64352 & 0.81709 & 1.0147 & 0.7845 & 0.7732\\ \hline 
2022 & 0.81013 & 0.62775 & 0.81029 & 1.0116 & 0.7749 & 0.7660\\ \hline 
\hline
\end{tabular}
%\end{adjustbox}
\label{tab:tax}
\end{table}

\begin{table}
\centering
\caption{Gini-index ($g$), Pietra-index ($p$) and Kolkata-index ($k$)  values for the  inequalities from movie income  datasets containing
year-wise  box office earnings (1999-2022) for Bollywood (India) movies, collected
from ref. \cite{13bbollywood}. For each year, the total collections  of
individual films are recorded. Only years with more than 10 released
movies are included to ensure reliable statistical representation. Here
$N$ represents the number of produced and released movies in each year.}
%\begin{adjustbox}{width=0.7\textwidth}
\begin{tabular}{|l|c|c|c|c|c|c|c|}
\hline
%\textbf{Year} & N & \textbf{$g$-index} & \textbf{$k$-index} & \textbf{$p$-index} & $p/(2k-1)$ & p/g & 2k-1/g  \\ \hline
\toprule
\adjustbox{angle=90}{Year} &
\adjustbox{angle=90}{Tot. No. of movies} &
\adjustbox{angle=90}{Gini index ($g$)} &
\adjustbox{angle=90}{Pietra index ($p$)} &
\adjustbox{angle=90}{Kolkata index ($k$)}&
\adjustbox{angle=90}{$p/(2k-1)$} &
\adjustbox{angle=90}{$p/g$} &
\adjustbox{angle=90}{$(2k-1)/g$}\\
\midrule
\hline
1999 & 52 & 0.41093 & 0.29326 & 0.64469 & 1.0134 & 0.7136 & 0.7042\\ \hline 
2000 & 51 & 0.41637 & 0.29554 & 0.64453 & 1.0224 & 0.7098 & 0.6942\\ \hline 
2001 & 51 & 0.55679 & 0.41486 & 0.70681 & 1.0030 & 0.7451 & 0.7429\\ \hline 
2002 & 51 & 0.36364 & 0.26156 & 0.63000 & 1.0060 & 0.7193 & 0.7150\\ \hline 
2003 & 52 & 0.45222 & 0.33932 & 0.66963 & 1.0002 & 0.7503 & 0.7502\\ \hline 
2004 & 51 & 0.44658 & 0.34524 & 0.67206 & 1.0033 & 0.7731 & 0.7706\\ \hline 
2005 & 52 & 0.38679 & 0.29895 & 0.64882 & 1.0044 & 0.7729 & 0.7695\\ \hline 
2006 & 50 & 0.52425 & 0.41003 & 0.70370 & 1.0065 & 0.7821 & 0.7771\\ \hline 
2007 & 51 & 0.49243 & 0.36600 & 0.68287 & 1.0007 & 0.7433 & 0.7427\\ \hline 
2008 & 51 & 0.49477 & 0.36377 & 0.68170 & 1.0010 & 0.7352 & 0.7345\\ \hline 
2009 & 64 & 0.61159 & 0.46545 & 0.72633 & 1.0283 & 0.7610 & 0.7401\\ \hline 
2010 & 140 & 0.77307 & 0.63066 & 0.80724 & 1.0263 & 0.8158 & 0.7949\\ \hline 
2011 & 123 & 0.78239 & 0.65696 & 0.81366 & 1.0472 & 0.8397 & 0.8018\\ \hline 
2012 & 133 & 0.78140 & 0.64864 & 0.81088 & 1.0432 & 0.8301 & 0.7957\\ \hline 
2013 & 137 & 0.76403 & 0.60528 & 0.79360 & 1.0308 & 0.7922 & 0.7686\\ \hline 
2014 & 146 & 0.80382 & 0.65775 & 0.81760 & 1.0355 & 0.8183 & 0.7902\\ \hline 
2015 & 165 & 0.79784 & 0.65836 & 0.81791 & 1.0355 & 0.8252 & 0.7969\\ \hline 
2016 & 216 & 0.82970 & 0.68791 & 0.83025 & 1.0415 & 0.8291 & 0.7961\\ \hline 
2017 & 252 & 0.85300 & 0.70442 & 0.84432 & 1.0229 & 0.8258 & 0.8073\\ \hline 
2018 & 219 & 0.84407 & 0.71848 & 0.84777 & 1.0330 & 0.8512 & 0.8240\\ \hline 
2019 & 244 & 0.84798 & 0.72439 & 0.85467 & 1.0212 & 0.8543 & 0.8365\\ \hline 
2020 & 76 & 0.86551 & 0.73448 & 0.85860 & 1.0241 & 0.8486 & 0.8286\\ \hline 
2021 & 50 & 0.79775 & 0.63359 & 0.81054 & 1.0201 & 0.7942 & 0.7785\\ \hline 
2022 & 165 & 0.84379 & 0.69608 & 0.84151 & 1.0191 & 0.8249 & 0.8095\\ \hline 
2023 & 227 & 0.89157 & 0.77006 & 0.87826 & 1.0179 & 0.8637 & 0.8485\\ \hline 
2024 & 230 & 0.89728 & 0.76409 & 0.87516 & 1.0184 & 0.8516 & 0.8362\\ \hline 
\end{tabular}
  % \end{adjustbox}
    \label{tab:movie}
\end{table}

\begin{table}
\centering
\caption{Gini-index ($g$),  Pietra-index ($p$) and Kolkata-index ($k$)
 for the citation inequalities among the papers published by each of
the Nobel Laureates (collected, for years 2020-2025, from Google Scholar
\cite{google-scholar}). Here  $h$ represents the Hirsch index \cite{Hirsch05} for the individual
scientists listed.}
%\begin{adjustbox}{width=1.0\textwidth}
\begin{tabular}{|l|c|l|c|c|c|c|c|c|c|c|c|}
\hline
\adjustbox{angle=90}{Categories} &
\adjustbox{angle=90}{Year} &
\adjustbox{angle=90}{Name of Nobel laureates} &
\adjustbox{angle=90}{Tot. no. of papers} &
\adjustbox{angle=90}{Tot. no. of citations} &
\adjustbox{angle=90}{Hirsch index ($h$)} &
\adjustbox{angle=90}{Gini index ($g$)} &
\adjustbox{angle=90}{Pietra index ($p$)} &
\adjustbox{angle=90}{Kolkata index ($k$)} &
\adjustbox{angle=90}{$p/(2k-1)$} &
\adjustbox{angle=90}{$p/g$} &
\adjustbox{angle=90}{$(2k-1)/g$} \\
\hline
\multirow{7}{*}{\adjustbox{angle=90}{Economics}} 
 & 2020 & Paul Milgrom & 398 & 118799 & 83 & 0.9123 & 0.7965 & 0.8941 & 1.0105 & 0.8731 & 0.8640 \\ %\cline{2-12}
 & 2020 & Robert Wilson & 259 & 34825 & 58 & 0.8799 & 0.7379 & 0.8685 & 1.0012 & 0.8386 & 0.8376 \\ %\cline{2-12}
 & 2021 & David Card & 695 & 103999 & 117 & 0.8957 & 0.7627 & 0.8785 & 1.0075 & 0.8515 & 0.8452 \\ %\cline{2-12}
 & 2021 & Guido Imbens &393 & 119042 & 102 & 0.8869 & 0.7522 & 0.8751 & 1.0025 & 0.8481 & 0.8459 \\ %\cline{2-12}
 & 2021 & Josh Angrist & 436 & 107740 & 92 & 0.9212 & 0.7891 & 0.8922 & 1.0060 & 0.8566 & 0.8515 \\ %\cline{2-12}
 & 2022 & Ben Bernanke & 717 & 147600 & 118 & 0.9210 & 0.7984 & 0.8984 & 1.0021 & 0.8669 & 0.8651 \\ %\cline{2-12}
 & 2022 & Philip H. Dybvig & 139 & 55939 & 41 & 0.9426 & 0.8339 & 0.9139 & 1.0073 & 0.8846 & 0.8782 \\ %\cline{2-12}
 & 2024 & Daron Acemoglu & 1358 & 277596 & 183 & 0.9167 & 0.7928 & 0.8939 & 1.0064 & 0.8649 & 0.8594 \\ %\cline{2-12}
 & 2024 & James Robinson & 845 & 131696 & 102 & 0.9465 & 0.8293 & 0.9146 & 1.0001 & 0.8761 & 0.8760 \\ %\cline{2-12}
 & 2024 & Simon Johnson & 947 & 95164 & 64 & 0.9693 & 0.8962 & 0.9477 & 1.0008 & 0.9246 & 0.9238\\ %\cline{2-12} 
 & 2025 & Joel Mokyr & 471 & 32139 & 71 & 0.9049 & 0.7548 & 0.8747 & 1.0071 & 0.8341 & 0.8282 \\ %\cline{2-12}
 & 2025 & Philippe Aghion & 647 & 146570 & 145 & 0.8733 & 0.7205 & 0.8586 & 1.0046 & 0.8250 & 0.8212 \\ \hline

\multirow{6}{*}{\adjustbox{angle=90}{Physics}}
 & 2020 & Roger Penrose & 639 & 101446 & 102 & 0.9022 & 0.7727 & 0.8859 & 1.0011 & 0.8565 & 0.8555 \\ %\cline{2-12}
 & 2021 & Giorgio Parisi & 1127 & 111103 & 136 & 0.8437 & 0.6694 & 0.8346 & 1.0002 & 0.7934 & 0.7933 \\ %\cline{2-12}
 & 2021 & Syukuro Manabe & 298 & 49325 & 89 & 0.8291 & 0.6687 & 0.8282 & 1.0186 & 0.8065 & 0.7918 \\ %\cline{2-12}
 & 2022 & John F. Clauser & 138 & 22474 & 33 & 0.9450 & 0.8298 & 0.9137 & 1.0029 & 0.8781 & 0.8756 \\ %\cline{2-12}
 & 2022 & Aspect Alain & 772 & 41611 & 76 & 0.9342 & 0.7984 & 0.8972 & 1.0050 & 0.8546 & 0.8503 \\ %\cline{2-12}
 & 2023 & Anne LHuillier & 514 & 36201 & 86 & 0.8470 & 0.6792 & 0.8375 & 1.0063 & 0.8018 & 0.7968 \\ %\cline{2-12}
 & 2023 & Ferenc Krausz & 1122 & 90527 & 130 & 0.8860 & 0.7335 & 0.8644 & 1.0064 & 0.8279 & 0.8226 \\ %\cline{2-12}
 & 2024 & Geoffrey Hinton & 751 & 975672 & 192 & 0.9440 & 0.8310 & 0.9155 & 1.0001 & 0.8803 & 0.8802 \\ %\cline{2-12}
 & 2024 & John Hopfield & 306 & 94961 & 95 & 0.8964 & 0.7381 & 0.8687 & 1.0009 & 0.8234 & 0.8227 \\ %\cline{2-12}
 & 2025 & John Clarke & 1081 & 50125 & 111 & 0.8501 & 0.6913 & 0.8409 & 1.0140 & 0.8132 & 0.8019 \\ %\cline{2-12}
 & 2025 & Michel Devoret & 754 & 78163 & 133 & 0.8586 & 0.7110 & 0.8478 & 1.0220 & 0.8281 & 0.8102 \\ \hline

\multirow{6}{*}{\adjustbox{angle=90}{Chemistry~~~~}}
 & 2020 & Emmanuelle Charpentier & 269 & 64602 & 65 & 0.9309 & 0.8030 & 0.9014 & 1.0002 & 0.8626 & 0.8624 \\ %\cline{2-12}
 & 2020 & Jennifer Doudna & 864 & 156101 & 164 & 0.8582 & 0.6868 & 0.8429 & 1.0014 & 0.8003 & 0.7992 \\ %\cline{2-12}
 & 2021 & Benjamin List & 337 & 49436 & 103 & 0.7482 & 0.5753 & 0.7868 & 1.0029 & 0.7690 & 0.7667 \\ %\cline{2-12}
 & 2021 & David MacMillan & 578 & 87442 & 134 & 0.8288 & 0.6704 & 0.8296 & 1.0171 & 0.8089 & 0.7952 \\ %\cline{2-12}
 & 2022 & Carolyn Bertozzi & 1021 & 103476 & 153 & 0.8081 & 0.6306 & 0.8132 & 1.0067 & 0.7804 & 0.7752 \\ %\cline{2-12}
 & 2022 & Morten Meldal & 412 & 32092 & 68 & 0.8261 & 0.6382 & 0.8165 & 1.0083 & 0.7726 & 0.7663 \\ %\cline{2-12}
 & 2023 & Louis Brus & 468 & 92751 & 122 & 0.8175 & 0.6606 & 0.8274 & 1.0088 & 0.8081 & 0.8010 \\ %\cline{2-12}
 & 2023 & Moungi G. Bawendi & 989 & 178538 & 189 & 0.8323 & 0.6666 & 0.8314 & 1.0057 & 0.8009 & 0.7964 \\ %\cline{2-12}
 & 2024 & Demis Hassabis & 171 & 238294 & 100 & 0.8530 & 0.6991 & 0.8492 & 1.0009 & 0.8196 & 0.8189 \\ %\cline{2-12}
 & 2024 & David Baker & 2690 & 201046 & 227 & 0.8431 & 0.6868 & 0.8399 & 1.0102 & 0.8146 & 0.8064 \\ %\cline{2-12}
 & 2025 & Omar M. Yaghi & 741 & 275500 & 197 & 0.8478 & 0.6895 & 0.8427 & 1.0061 & 0.8133 & 0.8084 \\ %\cline{2-12}
 & 2025 & Susumu Kitagawa & 1129 & 104460 & 146 & 0.7994 & 0.6226 & 0.8112 & 1.0002 & 0.7788 & 0.7787 \\ \hline

\multirow{6}{*}{\adjustbox{angle=90}{Medicine}}
 & 2020 & Michael Houghton & 546 & 61896 & 107 & 0.8429 & 0.6749 & 0.8366 & 1.0025 & 0.8007 & 0.7987 \\ %\cline{2-12}
% & 2021 & Ardem Patapoutian & 185 & 53407 & 89 & 0.7359 & 0.5643 & 0.7786 & 1.0129 & 0.7669 & 0.7572 \\ \cline{2-12}
 & 2022 & Svante Paabo & 587 & 151556 & 179 & 0.7805 & 0.6006 & 0.7989 & 1.0046 & 0.7696 & 0.7660 \\ %\cline{2-12}
 & 2023 & Katalin Karikó & 226 & 32576 & 66 & 0.8429 & 0.6874 & 0.8408 & 1.0084 & 0.8155 & 0.8087 \\ %\cline{2-12}
 & 2024 & Victor Ambros & 179 & 73309 & 72 & 0.8827 & 0.7315 & 0.8653 & 1.0011 & 0.8287 & 0.8278 \\ %\cline{2-12}
 & 2025 & Shimon Sakaguchi & 676 & 143255 & 136 & 0.8794 & 0.7323 & 0.8638 & 1.0065 & 0.8327 & 0.8274 \\ \hline

\end{tabular}
%\end{adjustbox}
\label{tab:nobel}
\end{table}
\end{appendices}

\end{document}